# Generation of spherical and cylindrical shock acoustic waves from optical breakdown in water, stimulated with femtosecond pulse

F.V.Potemkin[*], E.I. Mareev , A.A. Podshivalov and V.M.Gordienko

*Faculty of Physics and International Laser Center M.V. Lomonosov Moscow State University, Leninskie Gory, bld.1/62, 119991, Moscow, Russia*
*Corresponding author: potemkin@automationlabs.ru*

Using shadow photography technique we have observed shock acoustic wave from optical breakdown, excited in water by tightly focused Cr:Forsterite femtosecond laser beam, and have found two different regimes of shock wave generation by varying only the energy of laser pulse. At low energies a single spherical shock wave is generated from laser beam waist, and its radius tends to saturation with energy increasing. At higher energies long laser filament in water is fired, that leads to the cylindrical shock wave generation, which longitude increases logarithmically with laser pulse energy. From shadow pictures we estimated maximal velocity in front or shock wave of 2300±150m/s and pressure of 1.0±0.1 GPa

Laser driven shock waves in transparent solids have been of keen interest for the past few decades [1–9]. The shock waves are generated when the laser-induced plasma begins to expand with ultrasonic speed. Due to high temperatures ($T_e$~10eV) in plasma thin layer of vapor surrounding plasma is generated [2,5]. As far as laser-induced plasma continues to expand, cool and vanish this layer forms a cavitation bubble [2]. In the case when the femtosecond laser pulse is focused in the bulk of condensed matter the cavitation bubble and shock front shape replicates the shape of laser induced plasma [5]. Initially the shock front propagates with the ultrasound speed that is equal to bubble wall velocity [2]. Then the shock wave is separated from the cavitation bubble due to the large difference in velocities between the shock wave front and the bubble wall [2,5]. Initially femtosecond laser-induced shock waves can achieve TPa pressures in solids and GPa in liquids [5,6,10,11] Such high pressure enables a set of applications including hypervelocity launchers1, synthesis of new materials [4,7], high-temperature and high-pressure plasma fields production, three dimension microfluid chips fabrication [8], ion separation [7], chemical reaction control [12], new unusual phase transition [4,13] and a variety of medical therapies [2], such as mechanical optical clearing [9]. In some cases [8,9,12] it may be interesting to control the shape and energy of laser-driven shock wave. For example the array of closely spaced shock waves can create define pressure profile, than made the impact on biological tissue [9] more complicated and précised or can accelerate the process of three- dimensional microfluidic chips fabricating [8] adding the additional degrees of freedom.

One possible way of shock wave shaping control is using the filamentation of intense femtosecond laser pulse. It occurs owing to the dynamic balance between two major physical effects: Kerr self-focusing and defocusing effects in the electron plasma generated through the ionization process [14–16]. When femtosecond laser radiation with peak power higher than critical power of self-focusing propagates in the nonlinear medium with nonzero , thin extended filaments of radiation and accompanying plasma channels are formed [14,15]. In nonlinear focuses due to high localization of laser pulse energy optical breakdown is occurred forming shock wave and cavitation bubble [2,14]. Tight focusing of laser radiation into condensed matter leads to single continuous plasma channel is formed and it generates continuous cylindrical shock wave [5,14,15].

The key point of our research is how varying all the available in the experiment parameters of intense (~$10^{13}$ W/cm$^2$) femtosecond laser pulse propagating in strongly absorbing media with peak power much above critical power of self-focusing one could control the extreme state and shock wave characteristics. In our experiments with water cell the Cr:forsterite femtosecond laser pulse peak power was varied in range from $P_{cr}$ up to 50 $P_{cr}$. The creation of extreme state of strongly linear absorbing matter produced at a peak power much greater than the critical power of self-focusing for the given medium accompanied by the filament creation and its evolution on the nanosecond time scale are very attractive but poorly studied [14,15,17]. Incident laser energy which absorbs by the water molecule in linear regime goes to the excitation of the superposition oscillation of the last one and as a result the most part of laser energy does not reach the region of optical breakdown. Such specific interaction regime of laser radiation with matter may be very perspective in mentioned above mechanical optical clearing and could be implemented in it [9] because the most relevant laser radiation to solve this task lies in mid IR range due to deeper penetration lengths, low scattering by biological tissues, and possibilities of deep damaging with on-line controlled morphology of produced defects. However we will show we have extreme state formation and powerful shock wave generation in a bulk.

**Experimental setup.** In the experiments, Cr:forsterite femtosecond laser system (wavelength is 1240 nm, pulse duration full width at 1/e is about 140 fs, laser energy up to 150 μJ, intensity contrast is about 300, repetition rate is 10Hz) was employed. To observe the dynamics of the shock waves the shadow photography technique was applied. In this technique the shock waves and cavitation bubbles were observed on the CCD matrix (exposure relates to the time between laser pulses), due to the fact of probe pulse scattering on the refractive index modulations caused by mechanical effects. The experimental scheme is sketched at Fig.1.

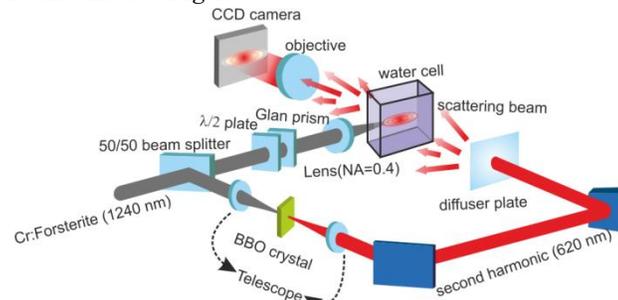

*Fig.1 Experimental setup. The incoming femtosecond laser pulse is splitted into two channels. The pump pulse is tightly focused into the water cell. The energy of pump pulse is varied by half-wave plate with fixed Glan prism. The probe pulse is transformed in its second harmonic in BBO crystal and scattered. Passing through water cell with shock wave modulated reflective index the probe beam is collected by the microscopic objective on the CCD matrix.*

Beam splitter divided initial beam into two beams with equalization in energy. The half-wave plate with Glan prism was used for continuous adjustment of the laser energy up to 150 µJ. The pump pulse was tightly focused by the lens (numerical aperture was 0.4, 3.3 and 4.6 mm focal lengths for different lenses, the diameter of the incident beam was slightly above the clear lens aperture) into the water cell, creating plasma, cavitation bubble and shock waves. The lenses were selected for observations of different plasma-induced regimes. Measuring at air conditions the focal spot diameter and the Rayleigh length for the lens with 3.3 mm focal length are about 4µm and 15µm respectively. Lenses were positioned close to the water cell. For the linear absorption measuring the defocused laser beam passage through 4mm water cell was used. It was established about 30% of energy passed through the cell. Therefore taking into account linear absorption about 0.9 cm-1 only 27% and 16% of incoming energy is delivered to the focal region in the case of lens with 3.3 and 4.6 mm focal length respectively. For the visualization on CCD matrix the probe pulse was transformed into the second harmonic in BBO crystal. Then the probe beam was scattered by diffuser plate for water cell uniform illumination. The scattered radiation was focused on CCD matrix with microscopic objective for region of interest observation. For the probe signal highlighting a band pass filter (620±10nm) was used. In tight focusing geometry constructed by using the lens with 3.3 mm focal length the filter was removed for plasma channel observation. Experiments were carried out both in collinear and non-collinear schemes. The angel between pump and probe pulse was equal to 45°.

**Results and discussion.** Examples of shadowgrams are sketched at Fig 2 and Fig.5. In order to improve images quality an averaging over ten images, subtracting the background, was performed. Using technique allowed us to observe the shock wave dynamics. It is well known, that there is a threshold of optical breakdown and shock wave formation can be served as an optical breakdown indicator [5]. Based on the shadowgrams the threshold of shock wave formation registration equals to 6±1µJ. This threshold can serve as an indicator of laser-induced optical breakdown in water [5]. Taking into account linear absorption the optical breakdown threshold equals to 1±0.2 µJ which is in a good agreement with the results obtained in similar focusing geometry by other methods both in water and other transparent dielectrics with correction for given wavelength [5,18–21]

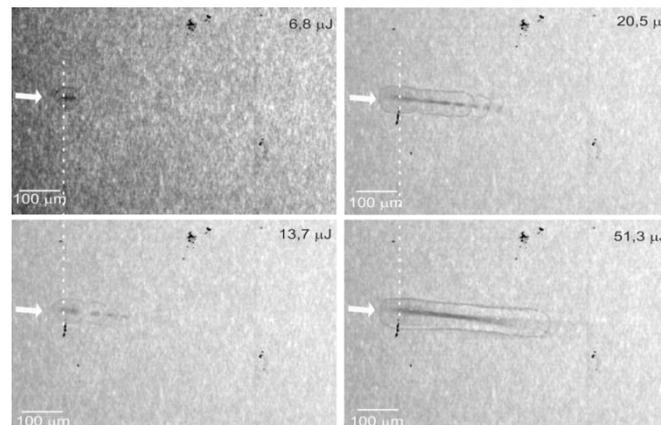

*Fig.2 The shadowgrams of 18.6 ns delay at different laser pulse energy for numerical aperture NA=0.4 and focal length f=4.6mm. Dashed line shows the center of initial plasma formation region. Arrow shows laser pulse direction. The dark regions in shadowgrams shows shock wave location and dark regions in the center of image shows cavitation region.*

The first experiments were carried out with not so tight focusing geometry (N.A. is equal to 0.4, lens focusing distance is equal to 4.6 mm at air conditions). Using this focusing regime the filamentation process is developed. The nonlinear focuses where intensity and therefore electron concentration is enough for optical breakdown becomes the centers for cavitation bubble and spherical shock waves generation [22]. Thus the spherical shock waves are isolated from each other, creating an array (Fig.2). The distance between two following shock waves is decreased. With the increase of energy the density of created plasma growth and additional shock waves are generated, forming a single cylindrical shock wave.

Using shadowgrams shock wave diameter can be measured, and D(t) and D(E) dependences can be restored (Fig. 3,4). The subsequent shock wave dynamics is fully described using hydrodynamic equations [5].

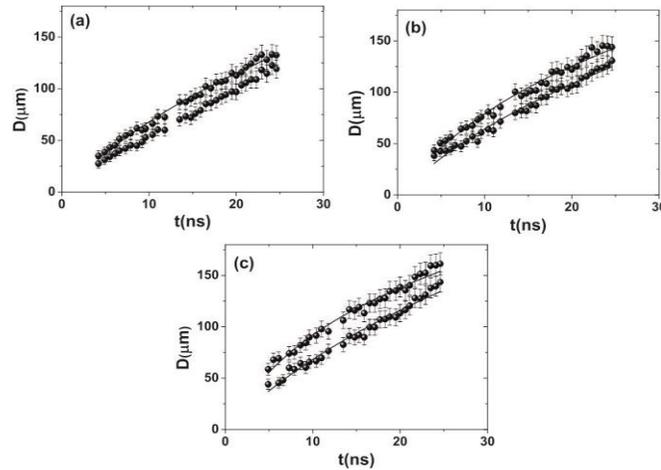

*Fig.3 Shock wave leading and trailing edge diameters D as a function of time t. Laser energies are equal to a) 10 µJ, b) 40 µJ and c) 120 µJ. Lines show approximation with shock velocity exponential decay.*

In the experiments we found, that shock wave diameter has an inverse quadratic attainment on saturation. It can be explained, using the fact, that there is a plateau in electron concentration dependence on laser pulse intensity when femtosecond laser radiation is tightly focused in the bulk of transparent media because the number of electrons in the region, where laser-matter interaction is taken place, is limited [23]. Thus the energy that can be transferred from laser radiation to plasma and then from plasma to each shock wave in the optical breakdown volume is limited. The shock wave energy has the quadratic dependence on its radius, due to mass flow conservation and its spherical symmetry [5].

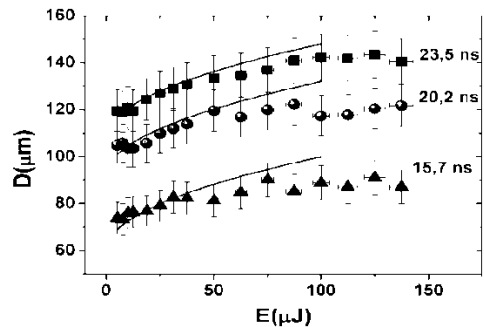

*Fig.4 Shock wave diameter D as a function of laser energy E. Time delays are equal to 15.7, 20.2 and 23.5 ns respectively. Lines show inverse quadratic dependence.*

Using the experimental dependence of the shock wave radius on time it is easy to estimate the shock wave front velocity. In the simplest case the speed decreases exponentially, reducing to sound speed[5]. To estimate the pressure at the shock front an empirical equation was used [2,5,24]. This is the upper limit of shock wave velocity value, because the decay constant is proportional to the velocity and such approximation is valid for times more than few nanoseconds. For incident laser energy of 130µJ shock wave front velocity is 2300±200m/s, the similar results can be found in [5,11]. The shock pressure equals to 240±30 MPa that is similar to results in papers [5,11]

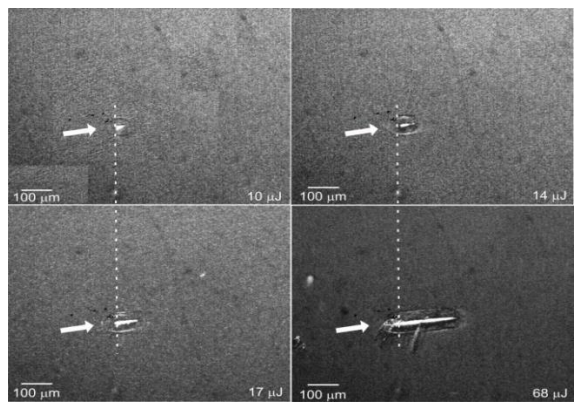

*Fig.5 The shadowgrams of 27ns delay at different laser pulse energy (NA=0.4; f=3.3 mm). Dashed line shows the center of initial plasma formation region. Arrow shows laser pulse direction. The dark regions in shadograms shows shock wave location and light regions in the center of image show the plasma channel location. For plasma channel observation a bandpass filter (620±10nm) was removed.*

Next we performed the experiments in tighter focusing regime and as a result in another plasma formation conditions. In this case single continuous plasma channel is observed. At low laser energies there is single spherical shock wave, which dynamics is described above. With the increase of laser energy the additional spherical shock waves are generated. And, as one can see from Fig.5, the cylindrical plasma channel is formed. Laser energy increasing leads to the transformation of the spherical shock waves array into cylindrical shock wave. Each point, where electron concentration is high enough, as in previous case, becomes a center of independent optical breakdown [21]. Thus laser energy growth led to optical breakdown location rising, and its moving toward the focus [5,25]. When electron concentration is higher than the shock wave formation threshold a new spherical shock wave is generated. The number of such shock waves is proportional to plasma channel length.

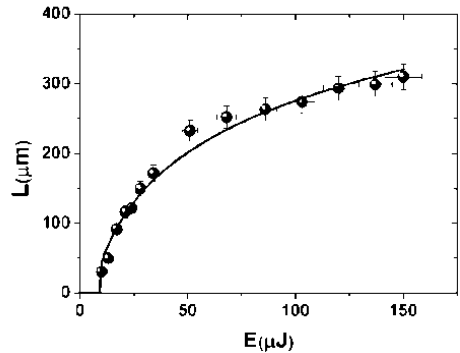

*Fig.6 Plasma channel length L as a function of laser energy E. Line corresponds to numeric calculations given by Keldysh approach for ionization.*

The length of the cylindrical region occupied by the shock wave equals to the sum of cavitation bubble region plus the shock wave diameter, because the center of the last spherical shock wave matches the center of the last cavitation bubble formation. Plasma channel location is practically undistinguishable from the cavitation bubble region, because the bubbles are the result of water evaporation by laser-induced plasma. Therefore namely plasma parameters determine shock wave pressure and its shape. Thus all mechanical effects depend on plasma parameters and in order to describe the shock wave dynamics we can focus on plasma channel behavior. Plasma channel length determines the shape of the cylindrical shock wave. To find its dependence on laser pulse energy using moving breakdown model together with Keldysh approach for ionization [26] a simple numerical simulation was carried out. In our approach we have taken into account that due to nanosecond time scale of shock processes the temporal dependence on laser pulse arrival can be vanished. Electron concentration at optical breakdown is set equal to electron concentration calculated for observable shock wave creation threshold. An excellent agreement with experiment is shown in Fig. 6, similar plasma channel length behavior in gases and solids one can find in papers [20,27,28].

**Summary.** For the first time a scheme for laser-driven plasma assisted shock wave shaping control by varying the femtosecond laser pulse energy and focusing parameters in a bulk of condensed matter is proposed and implemented. It was shown that despite the strong linear absorption of water at laser fundamental wavelength of 1.24 µm, the delivered energy from tightly focused laser beam to medium leads to filament production and extreme state creation in a bulk which as a result successfully goes to the generation of laser-induced shock wave with hypersonic speeds and pressures in the hundreds of kilo bars. It was found that tightly focusing femtosecond laser radiation into the water cell led to the spherical shock wave generation at the threshold energy. With laser pulse energy increasing multiple spherical shock waves were generated. Thus, depending on plasma formation regime (tight or loose focusing), they may form a single cylindrical shock wave or series of detached shock waves that with increasing laser energy also degenerate into a cylindrical shock wave. In plasma assisted regime the diameter of the cylindrical shock wave is limited, which is associated with saturation of the energy transferred from the plasma electrons to shock wave per unit volume. The length of the area occupied by laser-induced shock wave increases with energy of tightly focused femtosecond laser pulse as well as the corresponding plasma channel.

The authors thank Irina Zhvania for drawing the experimental setup. This research has been supported by the Russian Foundation for Basic Research (Project No. 14-02-00819 A) and partly by the M.V. Lomonosov Moscow State University Program of Development.

**References.**